\documentstyle[sprocl,epsf]{article}

\newcommand{\be}{ \begin{eqnarray}}
\newcommand{\ee}{\end{eqnarray}}
\newcommand{\beno}{ \begin{eqnarray*}}
\newcommand{\eeno}{\end{eqnarray*}}
\newcommand{\raf}[1]{(\ref{#1})}
\newcommand{\bold}[1]{\mbox{\boldmath $#1$}}    

\newcommand{\ie}{{\em i.e.}}                    
\newcommand{\q}{{\bf q}}                        
\newcommand{\beq}{\begin{equation}}
\newcommand{\eeq}{\end{equation}}
\newcommand{\beqar}{\begin{eqnarray}}
\newcommand{\eeqar}{\end{eqnarray}}
\newcommand{\bfig}{\begin{figure}}
\newcommand{\efig}{\end{figure}}
\newcommand{\DCC}{{\rm DCC}}                    

\begin{document}
\hfill LBNL-40176\\
\ \\
\begin{center}
{\small Proceedings, International Workshop on Astro Hadron Physics,\\
`Hadrons in Dense Matter',\\ APCTP, Seoul, Korea, October 1997}
\end{center}
\ \\
\ \\
\addtolength{\topmargin}{3cm}

\title{DILEPTONS FROM DISORIENTED CHIRAL CONDENSATES}

\author{VOLKER KOCH, J{\O}RGEN RANDRUP, XIN-NIAN WANG}

\address{Nuclear Science Division, Lawrence Berkeley National Laboratory,\\
University of California, Berkeley, CA 94720, USA}

\author{Y. KLUGER}

\address{Theoretical Division,
Los Alamos National Laboratory, Los Alamos, NM 87545, USA}


\maketitle\abstracts{
Disoriented chiral condensates are manifested as long 
wavelength pionic oscillations
and their interaction with the thermal environment can be a significant
source of dileptons. We calculate the yield of such dilepton production
within the linear sigma model and illustrate the basic features of the 
dilepton spectrum in a schematic model. We find that the dilepton yield with 
invariant mass near and below $2m_{\pi}$ due to the soft pion modes
can be up to two orders of magnitude larger than the corresponding 
equilibrium yield. We conclude with a discussion on how this enhancement 
can be detected by present dilepton experiments.
}

\section{Introduction}
\label{intro}

Recently much attention has been devoted to the phenomenon of so called
disoriented chiral condensates 
(DCC)\cite{BKT,rw93,GGP94,gm94,ahw,boy,kluger,kluger2,JR:PRL,ar89,bk92,bj92}. 
 The basic idea
behind the formation of a DCC is that after the restoration of chiral symmetry
in a given region of space-time the system may relax into a misaligned vacuum
with a non-vanishing expectation value of the pion field. 
The system will then eventually decay  into the normal vacuum by emitting 
soft and coherent
pions.  In the framework of ultrarelativistic heavy ion collisions a possible
way to generate a DCC state is a rapid cooling-like quenching
\cite{rw93,GGP94,gm94,ahw,boy,kluger,kluger2,JR:PRL} which leads to a
significant enhancement of low momentum pion modes.  The resulting occupation
numbers may then become large and lead to the emission of many pions in the
same isospin state.  In such an ideal scenario, the neutral pion fraction 
$f = <\pi_0>/<\pi>$
exhibits an anomalous distribution, $P(f)=1/2\sqrt{f}$, which has been
suggested as an experimental signal \cite{ar89,bk92,bj92}.  However, if several
separate domains are formed (\ie\ if the size of the system is large in
comparison with the correlation length), as may well occur in heavy-ion
collisions, the signal is correspondingly degraded and the distribution
approaches its normal form, an approximately normal distribution centered
around $f$=$1\over3$ \cite{Amado96,JR:NPA}.  More advanced methods of analysis
would then be needed, such as a the use of wavelets \cite{wave} or cumulative
moments \cite{minimax}.

The enhancement of
low-momentum pion modes, which is associated with the formation of DCC states,
should naturally lead to a strong enhancement in the dilepton channel. While
the
collective oscillation of the isospin current leads to an appreciable signal
only at very small invariant masses\cite{zhxw96}, the annihilation of the 
low-momentum
modes should result in a strong enhancement close to $M_{inv} \simeq 2 m_\pi$.
Moreover the signal should be confined to small momenta of the virtual 
photon since
only the lowest momentum pion modes are enhanced. This enhancement will be the
subject of my talk. It should be further noted that a similar enhancement 
has been predicted in  the photon channel  \cite{boy2}.
More details can be found in ref.\cite{KKRW97}.

\begin{figure}[htb]
\setlength{\epsfxsize=1.0\textwidth}
\centerline{\hspace{0.15\textwidth} \epsffile{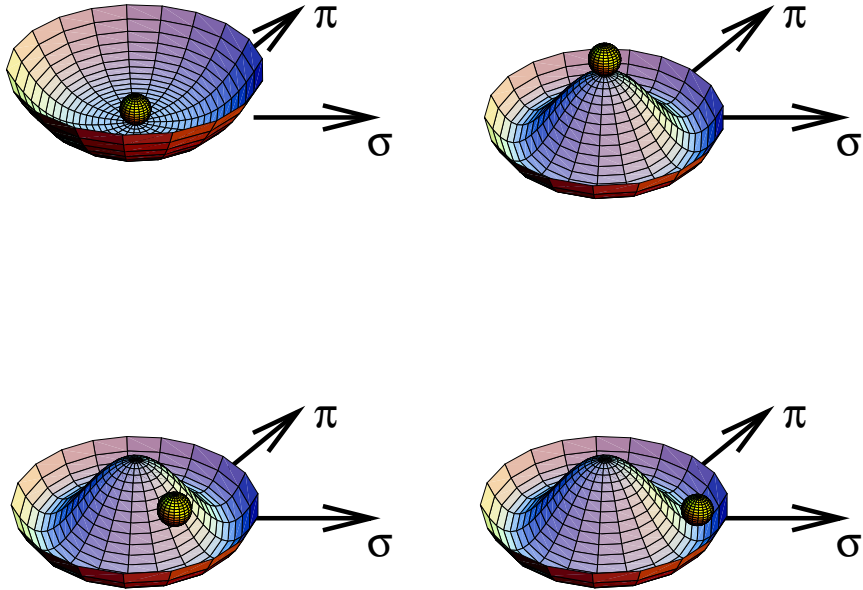}}
\caption{Schematic view of the quench scenario}
\label{potentials}
\end{figure}

\section{Schematic picture of a DCC}
Let us start by defining what we mean by a quench and by convincing ourselves
that such a quench indeed leads to an enhancement of low momentum pion modes.
The basic idea of a quench is illustrated in Figure \ref{potentials}:
Assume chiral symmetry has been  restored, i.e. the system sits in middle of 
the effective potential, $<\sigma> = 0$. After rapid cooling it might happen
that the system remains in the middle while the 
effective potential changes to a Mexican hat shape. This situation is highly 
unstable and the system wants to relax down into the rim of the Mexican hat. 
Since initially there is no prefered
direction in isospin, the system may roll down in any direction, 
$\sigma, \pi_1, \pi_2,  \pi_3$, where
$\pi_i$ denotes the Cartesian components of the pion field. Thus large 
amplitudes of the pion
field are generated before finally, due to explicit chiral symmetry breaking, 
the system eventually relaxes into the true vacuum, 
$<\sigma>=f_\pi$, $<\pi> = 0$. 

In order to see that the low momentum modes are enhanced, let us look at the
equations of motion of the pion field in the linear sigma model (here we ignore
the effect of thermal fluctuation for simplicity). 
\be
(\partial_t^2  - \nabla^2)\vec{\Phi} + \lambda 
(\vec{\Phi}^2 +  \sigma^2  -   v^2 ) \vec{\Phi} = 0 
\ee
with the general solution
\be
\Phi(t) = \Phi_0 \rm  e^{\pm i \omega_k t}
\ee
where
\be
\omega_k =  \sqrt{{m_\pi^*}^2 + k^2}
,\,\,\,\, {\rm and} \,\,\,\, { m_\pi^*}^2 \simeq 
\lambda ( <\sigma >^2 -  v^2 ) 
\ee

In the ground state, $<\sigma> = f_\pi$ and ${m_\pi^*} = m_\pi$. In the quench
scenario, on the other hand, $<\sigma> \simeq 0$ and ${m_\pi^*}^2 \simeq -
\lambda v^2 <0$ and hence for momenta smaller than the absolute value of the
pion mass, $k < |m_\pi^*|$, the frequency  $\omega_k$ turns out to be 
imaginary and consequently the modes grow exponentially.
\be
\Phi(t) \simeq \Phi_0 \rm  e^{+  |\omega_k| t}
\ee
Thus, only the modes with   $k < |m_\pi^*| \simeq 200 \, \rm MeV$ 
will be enhanced.
The different orientations of the DCC fields in isospin space result from the
the initial value of the pion field $\Phi_0$. Initially, before the quench
occurs, $\Phi_0$ is more or less randomly oriented in isospin space. Therefore,
depending on the direction  $\Phi_0$ points at the time of the quench different
isospin modes become amplified.

\begin{figure}[htb]
\setlength{\epsfxsize=0.65\textwidth}
\centerline{\hspace{0.15\textwidth} 
\epsffile{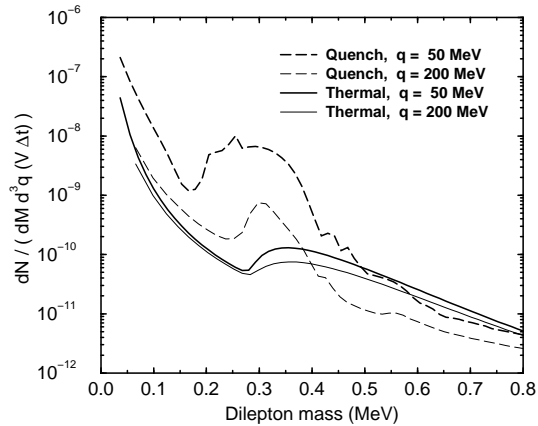}}
\caption{Dilepton invariant mass spectra for thermal (full lines) and quench
(dashed lines) initial conditions. Shown are the spectra for two different
values of the three dilepton three momentum $q$. The result has been obtained
in the 1/N mean-field approximation to the linear sigma model.}
\label{fig:quench_comp}
\end{figure}
\begin{figure}[htb]
\setlength{\epsfxsize=0.65\textwidth}
\centerline{\hspace{0.15\textwidth} 
\epsffile{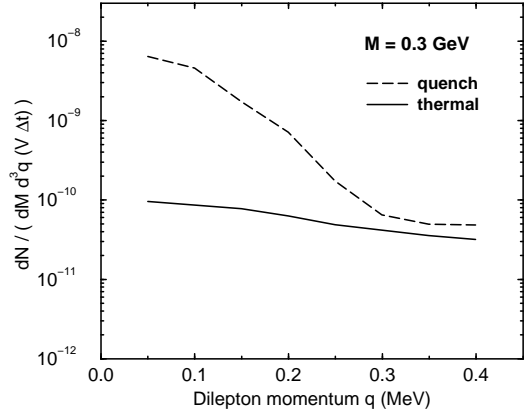}}
\caption{Momentum spectra for thermal (full lines) and quench
(dashed lines) initial conditions for dilepton pairs of invariant mass
$M= 300 \, \rm MeV$. The result has been obtained
in the 1/N mean-field approximation to the linear sigma model.} 
\label{fig:qplot}
\end{figure}
\begin{figure}[htb]
\setlength{\epsfxsize=0.65\textwidth}
\centerline{\hspace{0.15\textwidth}
\epsffile{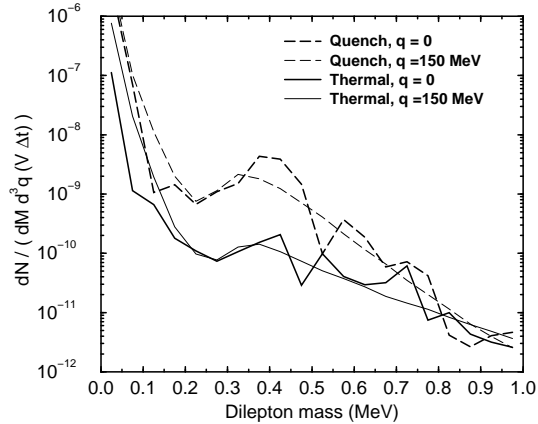}}
\caption{The invariant mass spectrum for both thermal initial conditions
(solid curves) and the corresponding quench scenario (dashed curves),
for either back-to-back dileptons having {\bf q}={\bf 0} (heavy curves)
and dileptons with a finite momentum of $q$=150~MeV(light curves).
The result has been obtained
in the semi-classical approximation to the linear sigma model.}
\label{JR:2}
\end{figure}
\begin{figure}[htb]
\setlength{\epsfxsize=0.65 \textwidth}
\centerline{\hspace{0.15\textwidth}
\epsffile{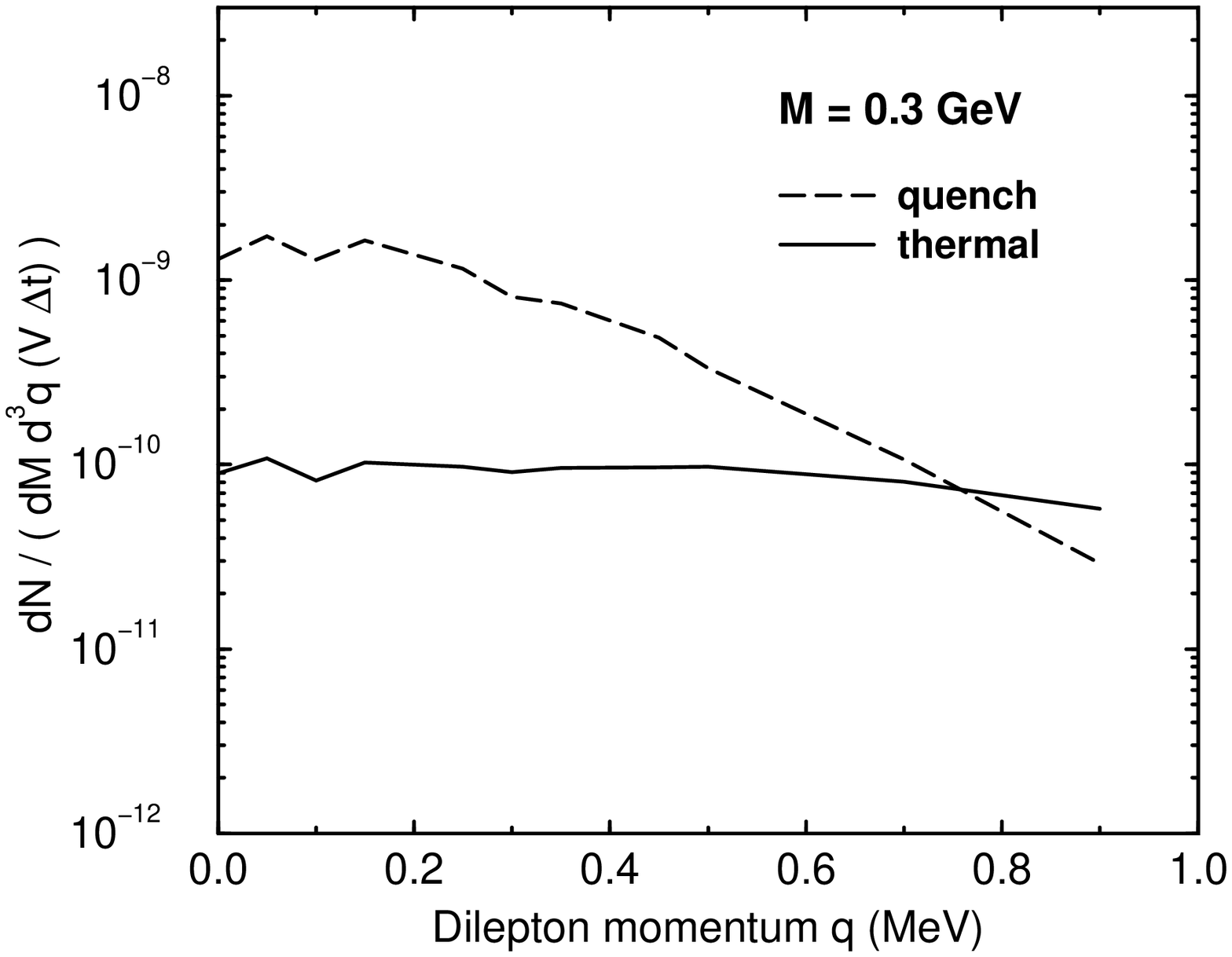}}
\caption{
The production rate $(d^4N/dM d^3q \,(V \Delta t))$
as a function of the magnitude of the dilepton momentum $\q$,
for dilepton masses near $M$=300~MeV. The result has been obtained
in the semi-classical approximation to the linear sigma model.}
\label{JR:3}
\end{figure}


 \section{Dilepton production}
 \label{dileptons}

 A detailed discussion on how to calculate the production of dileptons
from the dynamics of the sigma model can be found in ref. \cite{KKRW97}. 
We, therefore, restrict ourselves to the basic concepts and formulae.
 In general, in a non-equilibrium system, such as the dynamical
 evolution of DCC fields, the in and out states are not asymptotic 
 states and the density matrices describing the system
 are not diagonal (except in isospin and charge). In this paper,
 we will neglect quantum effects caused by the off-diagonal matrices 
 of the system in the calculation of the dilepton production.
 Therefore, the dilepton production yield can be given by the $S$ matrices
 of the electromagnetic transition between different states \cite{mclerran},
 \begin{eqnarray}\label{eq:yield0}
  \frac{dN_{\ell^+\ell^-}}{d^4q}&=&\frac{\alpha^2}{6\pi^3}\frac{B}{q^4}
    (q^\mu q^\nu-q^2g^{\mu\nu}) W_{\mu\nu}(q)\ , \nonumber \\
    W_{\mu\nu}(q)&=&\frac{1}{\cal Z}\int d^4x d^4y e^{-iq\cdot(x-y)}
    Tr [ \hat{\rho}\ \hat{j}_\mu(x)\  \hat{j}^\dagger_\nu(y) ]\ ,
  \label{dilepton_yield}
 \end{eqnarray}
 where we have summed over the final states, ${\cal Z}={\rm Tr}[\hat{\rho}]$, 
 and
 \begin{eqnarray}
   B=(1-\frac{4m^2_\ell}{q^2})^{1/2}(1+\frac{2m^2_\ell}{q^2})\ .
 \end{eqnarray}
 Since we are interested in the production of electron-positron pairs,
 we shall neglect the lepton mass $m_\ell$ in the following, implying $B=1$.

 For the actual calculation, we have modified the temporal 
integration boundaries from the
 asymptotic times $t=\pm\infty$ to finite initial and final times
 $t_i$ and $t_f$. This restriction leads to artificial `turn on / turn off' 
effects
 at low masses in the dilepton invariant mass spectrum as discussed in detail
 in \cite{KKRW97}. 
The dynamical input provided by the time-dependent solution of the linear sigma
model enters in eq. \raf{dilepton_yield} via the current-current correlation
function
\be
    Tr [ \hat{\rho}\ \hat{j}_\mu(x)\  \hat{j}^\dagger_\nu(y) ]. 
\ee
In the linear sigma model the electromagnetic current is given by
 \begin{equation}
   \hat{j}_\mu(x)=\frac{i}{2}[\hat{\pi}^\dagger(x)
   \stackrel{\leftrightarrow}{\partial}_\mu \hat{\pi}(x)-\hat{\pi}(x)
   \stackrel{\leftrightarrow}{\partial}_\mu \hat{\pi}^\dagger(x)]
  = \hat{\pi}_1(x) \partial_\mu \hat{\pi}_2(y) - 
     \hat{\pi}_2(x) \partial_\mu \hat{\pi}_1(y)\ , 
 \end{equation}
 where the complex charged pion field operators are related to the Cartesian 
 components by  
 \begin{equation}
   \hat{\pi}(x)=\frac{1}{\sqrt{2}}[\hat{\pi}_1(x)+i\hat{\pi}_2(x)]\ , \;\;
   \hat{\pi}^\dagger(x)=\frac{1}{\sqrt{2}}[\hat{\pi}_1(x)-i\hat{\pi}_2(x)] \ .
 \end{equation}
 We shall neglect the quadratic coupling in the gauged linear sigma model
 and the anomalous electromagnetic coupling of $\pi^0$, which all contribute
 to the dilepton production only to higher orders
 in the fine structure constant $\alpha=e^2/4\pi$.

We have calculated this correlation function in the $1/N$ mean-field 
approximation of the
linear sigma model by Kluger et al. \cite{kluger,kluger3} as well as in 
the semi-classical treatment of Randrup \cite{JR:PRD}.

In both model calculations, mean-field as well as semi-classical, we
have calculated the dilepton yield for two different initial conditions with
the
{\bf same} energy density.
\begin{enumerate}
\item Thermal initial conditions. These initial conditions are such that the
resulting system corresponds to a thermal ensemble of pions and sigmas. In this
case we have verified that the resulting dilepton yield agrees with that of a
pion gas.
\item Quench initial conditions. In this case the system is initialized at
small values for $\sigma$-field. Consequently
a considerable fraction of the thermal energy is taken up by the potential
(keeping the total energy density the same), leading to a cold initial
configuration well inside the unstable region. As a result a well developed 
DCC will be temporarily formed until the system finally returns to thermal 
equilibrium.
\end{enumerate}

In fig.\ref{fig:quench_comp} 
we show the resulting dilepton invariant mass spectra for thermal and
quench initial conditions for different three momenta of the virtual photon.
Clearly a strong enhancement (factor $\simeq 100 $) is seen around invariant
masses of $\simeq 2 m_\pi$. This enhancement is confined to small momenta, as
can be seen in fig. \ref{fig:qplot} where we plot the dilepton momentum
spectrum
for pairs having  invariant mass $M = 300 \, \rm MeV$. A similar behavior is 
found
in the semi-classical calculation (see figures \ref{JR:2} and \ref{JR:3}). Here
the enhancement is somewhat smaller which is partially due to the additional
mode mixing in the semi-classical treatment as well as the considerable larger
grid size, which produces a faster damping of the low-momentum modes.  

\section{A schematic model}
\label{schematic}

With the mean-field and semi-classical treatments,
the time-dependent fields are dynamically coupled to the quasi-particle modes
and the soft and hard modes of the fields are treated on an equal footing.
To understand the main features  of dilepton production due to the soft modes,
we shall now consider the dilepton production process
in a schematic model in which the soft and hard modes
are coupled only via the electromagnetic interaction
causing dilepton production.
The hard modes are then represented by a thermal gas of pions
having a specified temperature $T$,
while the soft modes will be evolved dynamically
according to the  one-dimensional Bjorken expansion scenario.
Hence the density operator can be factorized,
$\hat{\rho}=\hat{\rho}_{th} \otimes \hat{\rho}_{DCC}$,
where $\hat{\rho}_{th}$ is the normalized density operator for the thermal gas
and $\hat{\rho}_{DCC}$ is the normalized density operator for the soft modes,
referred to as the \DCC\ field.

The current-current correlator $\hat{W}_{\mu\nu}(q)$ can then be
decomposed according to how many thermal pions are involved
in the dilepton production process.
The first term represents the coherent emission of dileptons from \DCC\ fields
alone and involves no thermal pions,
\begin{equation}
  W^{(0)}_{\mu\nu}(q)\ =\ \int d^4x d^4y\
  \langle\hat{j}_\mu(x)\hat{j}^\dagger_\nu(y)\rangle_{DCC}\
        {\rm e}^{-iq\cdot(x-y)}\
  =\ \langle\hat{j}_\mu(q)\hat{j}^\dagger_\nu(q)\rangle_{DCC}\ ,
  \label{eq:yield1}
\end{equation}
where the \DCC\ expectation value is
$  \langle \cdots\rangle_{DCC}={\rm Tr}[\hat{\rho}_{DCC}\cdots ]$.

Contributions to the current-current correlator
from processes involving a thermal pion are
\begin{eqnarray}
  W^{(1)}_{\mu\nu}(q)&=&\int\frac{d^3k}{2\omega_k (2\pi)^3}
  \left\{ (2k_\mu+q_\mu)(2k_\nu+q_\nu) 
  [(1+n^+_k) \langle\hat{\pi}^\dagger(-q-k)\hat{\pi}(-q-k)\rangle_{DCC}
  \right. \nonumber \\
  &+&(1+n^-_k) \langle\hat{\pi}(q+k)\hat{\pi}^\dagger(q+k)\rangle_{DCC}]
  +(2k_\mu-q_\mu)(2k_\nu-q_\nu) \nonumber \\
  &\cdot& \left.
  [ n^+_k \langle\hat{\pi}(q-k)\hat{\pi}^\dagger(q-k)\rangle_{DCC}
  +n^-_k \langle\hat{\pi}^\dagger(k-q)\hat{\pi}(k-q)\rangle_{DCC}]
  \right\}, \label{eq:yield3}
\end{eqnarray}
The first two terms in the above equation correspond
to the emission of one pion together with a dilepton by the \DCC\ field.
The factor $1+n^\pm_k$ is a result of the Bose
enhancement in the final state.
The second  two terms proportional to $n^\pm_k$ represent the annihilation
or absorption of one thermal pion by the \DCC\ field.

From Eq.\ (\ref{eq:yield3}) one can readily see
how the momentum distribution of the \DCC\ field is imprinted
onto the dilepton spectrum.
Once the momentum of the dilepton is large
compared to the inverse of the \DCC\ domain size, 
the integral no longer has support from the 
Fourier transform of the pion field 
$\langle\hat{\pi}^\dagger(k\pm q)\hat{\pi}(k\pm q)\rangle_{DCC}$,
which restricts the contribution to low momenta and to a small window
in invariant mass.
For example, if one considers a classical field that oscillates
with a typical frequency of $\omega \simeq m_\pi$ and has a
Gaussian distribution of width $R_\perp$ in coordinate space,  
one can see from Eq.\ (\ref{eq:yield3}) that the dilepton yield 
will be concentrated around an invariant mass of $M\simeq 2 m_{\pi}$.
The width of this distribution will be of the order of $1/R_\perp$
in invariant mass as well as in the three-momentum $\bold{q}$. 

\begin{figure}[htb]
\setlength{\epsfxsize=0.65\textwidth}
\centerline{\hspace{1cm}
\epsffile{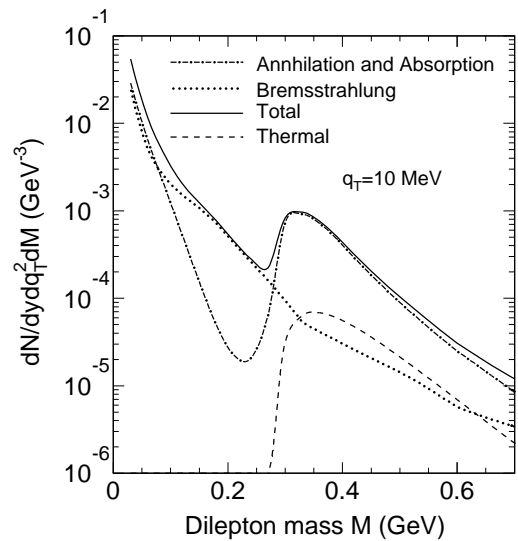}}
\caption{Dilepton spectrum from the \DCC\ field,
both bremsstrahlung (dotted),
annihilation and absorption (dot-dashed), and their sum (solid),
as well as the thermal emission (dashed). 
The initial temperature of the thermal environment is
$T_0$=145~MeV, while the characteristic energy density
carried by the \DCC\ field is $\epsilon_0$=58~MeV/fm$^3$.}
\label{fig:sch}
\end{figure}

Given this schematic model, 
it is possible to take approximate account of expansion
by subjecting the \DCC\ field to a boost-invariant Bjorken expansion.
For details of this calculation we refer to ref. \cite{KKRW97}.
The resulting dilepton invariant mass spectrum is shown in 
fig. \ref{fig:sch}
together with the corresponding average thermal rate.

First we find that the coherent emission rate  is 
negligible as compared to the incoherent emission.
Second, the incoherent dilepton emission is very significant
in comparison with the thermal production. The spectrum from
the pion annihilation and absorption by the \DCC\ field (dot-dashed)
has a structure manifest of two components in the second two terms
in Eq.~(\ref{eq:yield3}), depending on the energy flow.
In the contribution from the annihilation with the \DCC\ field,
energy flows out of the \DCC\ field, therefore the dilepton
spectrum has a threshold at $M=2m_\pi$, if the \DCC\ pion field
oscillate with a minimum frequency of $m_\pi$. If the energy flows
into the \DCC\ field, dileptons are then emitted when thermal pions
are absorbed by the \DCC\ field.
The dilepton spectrum from these absorption processes has no threshold
and dominates the annihilation and absorption spectrum at small invariant
masses $M<m_\pi$, as seen in fig.~\ref{fig:sch}.

Shown as the dotted line in fig.~\ref{fig:sch} is the dilepton spectrum
due to pion bremsstrahlung from the \DCC\ field.
These bremsstrahlung contributions result from transitions between different
modes of the DCC, which are a result of so called parametric resonances 
(for details see \cite{KKRW97}). 

In fig.~\ref{fig:sch} is also shown the sum of the different contributions
to dilepton spectrum from the \DCC\ field (solid curve)
as well as the contribution from annihilation in the cooling pion gas.
In general, one can see that the incoherent dilepton production
below and near $2m_\pi$ threshold region is significantly larger
than the thermal production.
As we already pointed out,
because of the finite spatial size of \DCC\ field in the transverse direction,
dileptons from the \DCC\ field exhibit a much faster decrease
with the transverse momentum $q_\perp$ than those due to thermal production. 

\section{Observational aspects}
So far we have established that the presence of DCC states leads to a strong
enhancement of the dilepton production at invariant masses close to $M \simeq 2
m_\pi$. We have also shown that this enhancement is confined to small momenta
of the virtual photons $p < 300 \, \rm MeV$. 
The question  of course  remains to which
extent this rather unique signal can be observed in an actual experiment, where
other, more conventional sources may dominate the dilepton measurement. For the
mass range under consideration the Dalitz decay of the $\eta$ meson should be
the most dominant background. Furthermore certain acceptance cuts need to be
applied in order to minimize the combinatorial background from false pairs.  In
order to estimate these various backgrounds, we have used a recent transport
calculation for dilepton production for Pb+Pb collision at 156 AGeV
\cite{KS96}. This calculation contains among others the $\eta$-Dalitz decay
channel as well as pion annihilation.  In order to calculate the contribution
from the pion annihilation on DCC states we have extracted from the above
mean-field calculation the ratio of quench to thermal dilepton production.
\be
R(M,p;\,M_{ref}=350 {\rm MeV}) = 
\frac{Y_{quench}(M,p)}{Y_{thermal}(M_{ref}=350 {\rm MeV},p)}
\label{ratio}
\ee
where $Y_{quench}$, $Y_{thermal}$ stand for the dilepton yield for given 
M and p. $M_{ref}$ is the value of the invariant mass for which the matching
between the two calculations is done for all $p_t$.
Assuming, that the dilepton production in the heavy ion 
collisions is more or less
thermal we can estimate the dilepton yield from DCC states in a Pb+Pb
collision by multiplying the dilepton yield from pion 
annihilation as obtained from the transport calculation 
with the above ratio \raf{ratio}
\be
Y_{DCC}(M,p)  = Y_{transp.}(M_{ref}=350 {\rm MeV},\,p) \, R(M,p;\,M_{ref}=350
{\rm MeV})
\ee
This of course ignores any non-thermal contribution as well as possible
temperature dependences of the enhancement factor. 

\begin{figure}[htb]
\setlength{\epsfxsize=1.0\textwidth}
\centerline{\hspace{0.15\textwidth}\epsffile{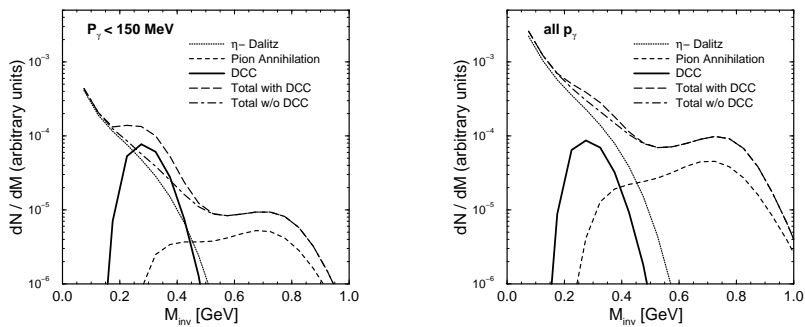}}
\caption{Dilepton invariant mass spectrum for Pb+Pb collisions at 158 AGeV as
measured by a {\it perfect} detector. Shown are only the channel relevant for
the discussion, but additional channel are taken into account in the total yield
(see \protect\cite{KS96}).}
\label{exp_per}
\end{figure}

In fig.  \ref{exp_per} we show the resulting dilepton spectra assuming
perfect detector acceptance (i.e. no acceptance cuts). In the left 
panel we show the spectrum for all dileptons with transverse momentum
smaller than 150 MeV. The right panel shows the dilepton mass spectrum
for all transverse momenta. Clearly, the signal is at least as large as
the competing $\eta$-Dalitz decay and thus should be observable
especially if the statistics allow to look at dilepton momenta below 150
MeV. 

\begin{figure}[htb]
\setlength{\epsfxsize=1.0\textwidth}
\centerline{\hspace{0.15\textwidth}\epsffile{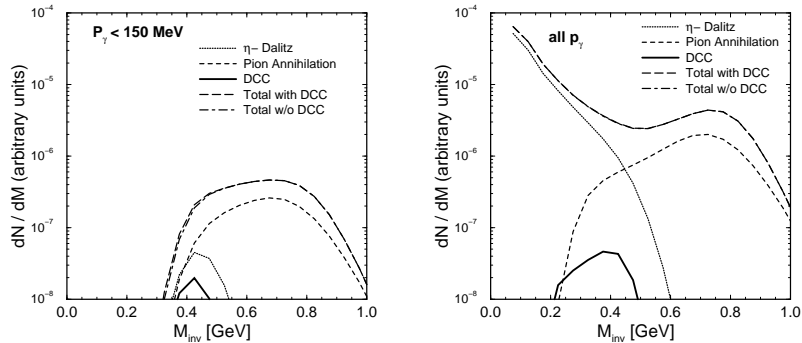}}
\caption{Same as Fig.~\protect\ref{exp_per}, but now the CERES acceptance have
been applied.}
\label{exp_ceres}
\end{figure}

In fig. \ref{exp_ceres} we have applied the cuts used by the CERES
collaboration \cite{CERES}. The most important cut is that on the
transverse momentum of each individual lepton. Only pairs where
each of the leptons has  $p_t > 200 \, \rm MeV$ are accepted. In this case, as
it is clear from the figure, our proposed signal is not visible.

\begin{figure}[htb]
\setlength{\epsfxsize=1.0\textwidth}
\centerline{\hspace{0.15\textwidth}\epsffile{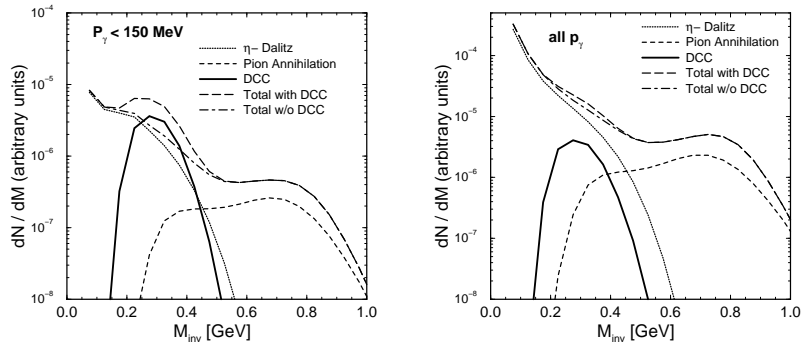}}
\caption{Same as Fig.~\protect\ref{exp_ceres}, but the CERES momentum cuts have
been reduced to 60 MeV.}
\label{exp_new}
\end{figure}

However, there might be a possibility that for small invariant masses 
CERES can relax the
transverse momentum cut values as low as 60 MeV \cite{axel}. In this
case the signal again should be visible as shown in fig.
\ref{exp_new}.
Of course our estimate started from the assumption that in each event
the conditions for a quench are satisfied. This might be somewhat
optimistic. However, if the system equilibrates in the high temperature
phase and generates sufficient rapid expansion this assumption should be
fulfilled to a large extent. Furthermore, we have not included any
dissipative processes which could destroy the DCC-states before the
system has frozen out. Consequently if the lifetime of a DCC turns out to
be considerably shorter than that of the system, our results have to be
reduced accordingly.

\section{Summary}

We have calculated the production of dileptons from disoriented chiral
condensates using a quantum mean-field as well as a semi-classical 
treatment for
the time evolution in the linear sigma model. We have compared the
dilepton spectra obtained when using so called quench initial conditions, 
which lead to a strong enhancement of long wave length pion modes (DCC), with
those obtained from thermal initial conditions. Compared to the thermal
spectrum the quench initial conditions lead to a strong enhancement (factor
20 - 100 depending on the model) at an invariant mass of about $M
\simeq 2
m_\pi$. This enhancement is confined to dilepton momenta  of 
$q \leq 300 - 500 \, \rm MeV$ 
and also rather narrow in invariant mass. 
Within a schematic model, 
we could also address the more realistic scenario of
a longitudinal expansion. We find that the above enhancement 
remains also in this case. 

Both dynamical solutions, which are based on distinct approximations to
the linear sigma model, predict qualitatively the same enhancement:
A large bump at $M \simeq 2 m_\pi$
as well as an enhancement at low invariant masses.
Overall,
the quantum mean-field model seems to yield a larger enhancement
than the semi-classical treatment.
This quantitative difference may be due to several factors.
First, the Bose enhancement factors are absent in the semi-classical treatment.
Second, the semi-classical treatment incorporates the mode mixing
resulting from the non-linear interaction and this mechanism
tends to reduce the number of pions in the DCC state.
Both of these features lead to a somewhat smaller signal.

As far as experimental observation of this enhancement is concerned, 
we have carried out a rough estimate of the expected signal for a Pb+Pb
collision at SPS energies. We have shown that the signal would be strong
enough to stand out from the main background, the Dalitz decay of the
$\eta$-meson. We furthermore have applied the acceptance cuts of the
CERES experiment. With the present cuts of $p_t > 200 \, \rm MeV$ the
signal would not be visible by the CERES detector. However, if these
cuts can be relaxed to $p_t > 60 \, \rm MeV$, the signal should be observable,
provided, of course that these DCC states are ever being formed in
SPS-energy heavy ion collisions.

\section*{Acknowledgments}
This work was supported by the Director, Office of Energy Research,
Office of High Energy and Nuclear Physics, Divisions of High Energy Physics
and Nuclear Physics of the U.S. Department of Energy under Contract
No. DE-AC03-76SF00098.


\end{document}